\documentclass[a4paper,12pt]{article}
\usepackage[english]{babel}
\usepackage[a4paper, left=20mm,right=20mm,top=20mm,bottom=20mm]{geometry}
\usepackage[normalem]{ulem}
\usepackage[dvips]{graphicx}
\usepackage{color}
\usepackage{setspace}
\usepackage{amsmath}
\usepackage{amssymb}
\usepackage{amsfonts}
\usepackage{indentfirst}
\usepackage{bbold}
\usepackage{cite}

\usepackage{authblk}

\title{Nonlinear dynamics of weakly dissipative optomechanical systems}

\author[1]{Thales Figueiredo Roque}

\author[1,2]{Florian Marquardt}

\affil[1]{Max Planck Institute for the Science of Light, Staudtstra{\ss}e 2, 91058 Erlangen, Germany}
\affil[2]{Institute for Theoretical Physics, Department of Physics, University of Erlangen-N{\"u}rnberg, Staudtstra{\ss}e 7, 91058 Erlangen, Germany}

\author[3]{Oleg M. Yevtushenko}

\affil[3]{Ludwig-Maximilians-Universit{\"a}t, Arnold Sommerfeld Center, and \protect\\
                Center for Nano-Science, Munich, DE-80333, Germany}

\makeatother

\begin{document}
\maketitle
\begin{abstract}
Optomechanical systems attract a lot of attention because they provide
a novel platform for quantum measurements, transduction, hybrid systems,
and fundamental studies of quantum physics. Their classical nonlinear
dynamics is surprisingly rich and so far remains underexplored. Works
devoted to this subject have typically focussed on dissipation constants
which are substantially larger than those encountered in current experiments,
such that the nonlinear dynamics of weakly dissipative optomechanical
systems is almost uncharted waters. In this work, we fill this
gap and investigate the regular and chaotic dynamics in this important
regime. To analyze the dynamical attractors, we have extended the
``Generalized Alignment Index'' method to dissipative systems. We
show that, even when chaotic motion is absent, the dynamics in the
weakly dissipative regime is extremely sensitive to initial conditions.
We argue that reducing dissipation allows chaotic dynamics to appear
at a substantially smaller driving strength and enables various routes
to chaos. We identify three generic features in weakly dissipative classical
optomechanical nonlinear dynamics: the Neimark-Sacker bifurcation
between limit cycles and limit tori (leading to a comb of sidebands
in the spectrum), the quasiperiodic route to chaos, and the existence of transient chaos.
\end{abstract}

\section{Introduction}

Cavity optomechanics \cite{aspelmeyer2014} aims to explore and exploit
the interaction between radiation fields and mechanical vibrations,
with important applications ranging from sensitive measurements to
quantum communication. The foundations for this research field were
established already at the end of the 60s, when the classical effects
of radiation on the motion of a test mass were studied in the context
of precision measurements \cite{braginsky1967,braginsky1970}. 
For an extended review we refer the reader to Ref.~\cite{aspelmeyer2014}.
In the past few years, a range of impressive achievements has been observed,
which includes topological transport in optomechanical arrays~\cite{schmidt2015, peano2016},
the engineering of nonreciprocal interactions
~\cite{bernier2017, fang2017, peterson2017, barzanjeh2017, ruesink2018, xu2019},
the generation of single phonon states using optical control~\cite{hong2017}, 
the generation of mechanical squeezed states~\cite{wollman2015},
measurement-based quantum control of mechanical motion~\cite{rossi2018},
conversion of quantum information to mechanical motion~\cite{reed2017}, 
conversion between light in the microwave and optical range~\cite{andrews2014},
single photon frequency shifters \cite{fan2016},
force measurements using cold-atom optomechanics~\cite{schreppler2014}, 
and the use of unconventional mechanical modes, like high frequency
bulk modes of crystals~\cite{renninger2018}, multilayer graphene~\cite{singh2014},
and the modes of superfluid helium~\cite{kashkanova2017}. 

Classical nonlinear optomechanics is relevant in the case of highly
populated optical and mechanical modes. Though it attracted slightly
less attention during the initial evolution of modern cavity optomechanics,
a number of significant theoretical studies have been devoted to understanding
the structure of the phase space, including limit cycles and multistability \cite{marquardt2006, ludwig2008,
loerch2014, wurl2016, schulz2016}, and 
chaotic dynamics \cite{bakemeier2015, djorwe2018}. Experimental studies have been
relatively rare, but important phenomena have already been observed, including
limit cycles \cite{kippenberg2005,metzger2008}, period doubling and
chaos \cite{carmon2007, monifi2016, mwang2016, wu2017, navarro-urrios2017, jin2017},
the predicted multistable attractor diagram \cite{krause2015, buters2015}
which is characteristic for optomechanical systems, as well as further
aspects \cite{leijssen2017, doolin2014}. More
recent studies have exploited the coupling of several OM limit cycle
oscillators to explore OM synchronization dynamics. OM synchronization was first predicted
theoretically in \cite{heinrich2011}, then observed
experimentally for few-mode systems \cite{zhang2012, bagheri2013, zhang2015, colombano2019},
and analyzed in subsequent theoretical studies of large-scale lattice
dynamics \cite{holmes2012, lauter2015, weiss2016, lauter2017}.

Many theoretical works on nonlinear classical OM dynamics have considered
mainly systems operating outside the so-called resolved sideband regime.
This means that the optical dissipation is assumed to be of the same
order or larger than the mechanical frequency. At the same time, the
mechanical quality factor is often assumed relatively small, of the
order of $O(10^{3})$. For instance, the authors of Ref.~\cite{bakemeier2015}
have shown that limit cycles in such strongly dissipative OM systems
undergo a period doubling cascade and become chaotic attractors.

On the other hand, most state-of-the-art experiments reach the resolved
sideband regime and deal with substantially larger mechanical quality
factors, ranging from $10^{4}$ to $10^{9}$ (cf. Figs.~11 and 10
in Ref.~\cite{aspelmeyer2014}). These experiments raise a natural
question: do such weakly dissipative systems show something qualitatively
new in their classical dynamics? The straightforward guess is: yes,
because nonlinear phenomena are expected to be enhanced with decreasing
dissipation. For instance, the Hopf bifurcation \cite{ludwig2008},
at which an equilibrium point of the dynamics becomes unstable and
a limit cycle emerges, has a clear dependence on the dissipation constants.
The smaller the dissipation constants, the weaker the laser pumping 
needed to observe the Hopf bifurcation. Bistability, which is another
nonlinear phenomenon, follows the same rule. Of course, the possible
types of attractors are also very sensitive to the dissipation strength.
One could take one step further and ask whether the chaotic OM dynamics
is enhanced as well and acquires new features in the resolved sideband
regime.

In this work we investigate the nonlinear dynamics of weakly dissipative
OM systems. Weakly dissipative in this context is the same as sideband resolved,
meaning that the optical dissipation is much smaller than the mechanical frequency.
Firstly, we are interested in performing a classification
of attractors: whether they are chaotic or regular, what is their
dimensionality, etc. We show that the weakly dissipative regime is
much more complex and nontrivial than the strongly dissipative one.
In particular, the OM dynamics becomes very sensitive to the initial
conditions in the resolved sideband regime, which represents the first
substantial difference between the strongly and weakly dissipative
cases.

This sensitivity to initial conditions (as well as the long relaxation
times) makes the study of weakly dissipative OM systems computationally
very challenging. To overcome this problem, we suggest a new approach
to classify the attractors and to detect dynamical chaos. It is based
on the $GALI$ (Generalized ALignment Index) method \cite{skokos2001,skokos2004,skokos2007}
and has several advantages. Besides being significantly faster than
commonly used methods based on the calculation of the maximal Lyapunov
exponent, the modified $GALI$ method provides an efficient tool to
learn the dimensionality of the attractors. This has allowed us to
explore the OM attractors in a large range of parameters and to reveal
important phenomena which are well-known in nonlinear science but
have been overlooked so far in optomechanics. They include transient
classical chaos, quasiperiodic orbits, and routes to chaos beyond
the period doubling.

The rest of this paper is organized as follows: In Sect.~\ref{WeaklyDissOM},
we introduce the equations of motion of an OM system and discuss the
basic differences between the strongly and weakly dissipative regimes.
Sect.~\ref{GALI-sect} is devoted to the $GALI$ method and its extension
to the analysis of \emph{dissipative} nonlinear dynamics. We use this
method and our numerical simulations to present a diagram that illustrates
various regular and chaotic weakly dissipative dynamical regimes
in Sect.~\ref{GALI-OM}. In particular, we identify two generic features
that will become important in the exploration of nonlinear optomechanics:
a Neimark-Sacker bifurcation between limit cycles and limit tori (leading
to a comb of sidebands in the spectrum) and the existence of transient
chaos. In Sect.~\ref{ExpRel}, we discuss the experimental relevance
of our results. Finally, Sect.~\ref{Concl} contains our conclusions.

\section{Classical dynamics of a weakly dissipative optomechanical system
\label{WeaklyDissOM}}

\subsection{Equations of motion}

\begin{figure}[!ht]
	\centering
	\includegraphics[height=0.3\linewidth]{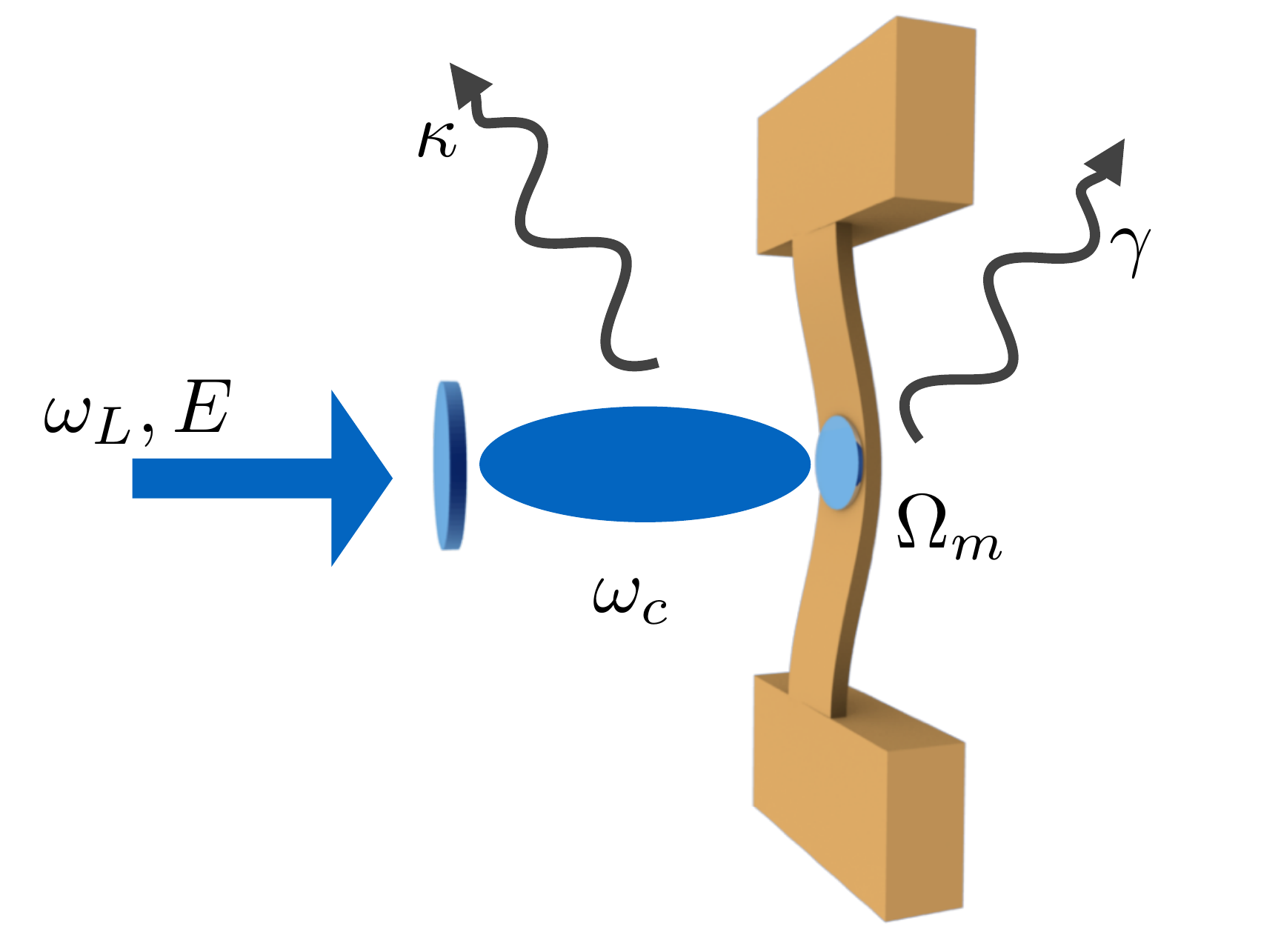}
	\caption{Representation of a generic OM system: an optical cavity with a movable
	mirror driven by an external laser.}
	\label{omsystem} 
\end{figure}

The classical dynamics of an OM system with one optical mode and one
mechanical mode (sometimes referred to as the optical cavity and the
mechanical oscillator, respectively) is described by the following
equations of motion \cite{aspelmeyer2014}: 
\begin{align}
	\frac{d}{dt}a &= (i\Delta-\kappa/2)a+ig_{0}a(b+b^{\ast})+E,
	\label{oc1}\\
	\frac{d}{dt}b &= (-i\Omega_{m}-\gamma/2)b+ig_{0}|a|^{2}.
	\label{oc2}
\end{align}
Here $b=(q+ip)/\sqrt{2}$, with $q$ and $p$ being the dimensionless
position and momentum of the mechanical oscillator, and $a$ is
the suitably normalized complex amplitude of the electric field inside
the cavity ($\left|a\right|^{2}$ and $\left|b\right|^{2}$ are the
photon and phonon number, respectively). The mechanical (optical)
mode has frequency $\Omega_{m}$ ($\omega_{c}$) and dissipation constant
$\gamma$ ($\kappa$). The optical mode is pumped by an external laser
with frequency $\omega_{L}$ and amplitude $E$; $\Delta=\omega_{L}-\omega_{c}$
denotes the detuning between the laser frequency and the cavity frequency;
$g_{0}$ is the bare optomechanical coupling constant. We note
that $E$ here is normalized such that $E^{2}/\kappa$ is the rate
of photons impinging on the cavity. The typical representation of an OM
system is shown in Fig.~\ref{omsystem}. As usual, we work in a reference
frame which rotates at the laser frequency%
\footnote{If $a$ is the complex amplitude of the electric field inside the
cavity, its counterpart in the lab frame reads $a_{{\rm lab}}=ae^{-i\omega_{L}t}$.}.
Eqs.~(\ref{oc1},\ref{oc2}) assume that quantum fluctuations can
be neglected, i.e. the dynamics is governed by highly populated optical
and mechanical states. These coupled equations have been employed
to describe countless experiments to high precision, both in the linearized
regime but also in the fully nonlinear regime of interest here.

Further numerical study requires to rewrite Eqs.~(\ref{oc1},\ref{oc2}) in a
dimensionless form. This can be done by defining rescaled
variables $\alpha=a\Omega_{m}/2E$ and $\beta=g_{0}b/\Omega_{m}$,
from which we obtain the following equations: 
\begin{align}
	\frac{d}{d\tau}\alpha &= \left( i\frac{\Delta}{\Omega_{m}} - \frac{\kappa}{2\Omega_{m}} \right) \alpha
	+ i\alpha(\beta + \beta^{\ast})+\frac{1}{2},
	\label{oc3}\\
	\frac{d}{d\tau}\beta &= \left( -i-\frac{\gamma}{2\Omega_{m}} \right) \beta + i\frac{P}{2}|\alpha|^{2},
	\label{oc4}
\end{align}
where $\tau=\Omega_{m}t$ and $P=8g_{0}^{2}E^{2}/\Omega_{m}^{4}$.
Note that there are fewer parameters in the rescaled Eqs.~(\ref{oc3},\ref{oc4})
than in the original Eqs.~(\ref{oc1},\ref{oc2}). This means the qualitative
features of the dynamics will only depend on four dimensionless combinations
of the original physical parameters: dimensionless power $P$, normalized
detuning $\Delta/\Omega_{m}$, normalized cavity decay $\kappa/\Omega_{m}$,
and mechanical dissipation $\gamma/\Omega_{m}$. For a more extended
discussion of the essential dimensionless parameters affecting classical
or quantum OM dynamics, we refer the reader to Refs.~\cite{ludwig2008,weiss2016}.

The parameter $P$ is a dimensionless measure of the laser input power,
which also includes the strength of the optomechanical interaction.
$P$ can be related to the standard measure of coupling strength vs.
dissipation, the so-called OM cooperativity $C=4g_{0}^{2}n_{c}/\gamma\kappa$.
Here $n_{c}$ is the mean number of photons stored in the optical
cavity. For our purposes the cooperativity is still slightly inconvenient,
since $n_{c}$ depends on the detuning (at fixed drive power). For
that reason, we rather introduce the \emph{maximum} cooperativity
$\tilde{C}=4g_{0}^{2}n_{0}/\gamma\kappa$, where $n_{0}=4E^{2}/\kappa^{2}$
is the number of photons in the \emph{resonantly} pumped optical cavity
in the absence of the optomechanical interaction. $P$ is then proportional
to the maximum cooperativity as follows:
\begin{equation}
	P = \kappa^{3}\gamma\tilde{C}/2\Omega_{m}^{4}\,.
	\label{oc5}
\end{equation}
This relation will be useful for comparison with experimental parameters.

\subsection{Fixed points}

Let us start our study of the dynamics with the analysis of the fixed points
of the system. Fixed points are points in
the phase space which are invariant under time evolution: if we take a fixed point as initial condition
of the system, the system stays on the fixed point forever. The analysis of trajectories whose
initial conditions are arbitrarily close to the fixed point allows one to classify the 
fixed point as stable, unstable, or hyperbolic. If any such trajectory is attracted to (repelled from)
the fixed point, the fixed point is stable (unstable). If some trajectories are attracted to the fixed point,
while other trajectories are repelled from it, the fixed point is hyperbolic. Stable fixed points are the
simplest attractors of a dynamical system.

\begin{figure}[!ht]
	\minipage{0.55\textwidth}
	\includegraphics[width=1\linewidth]{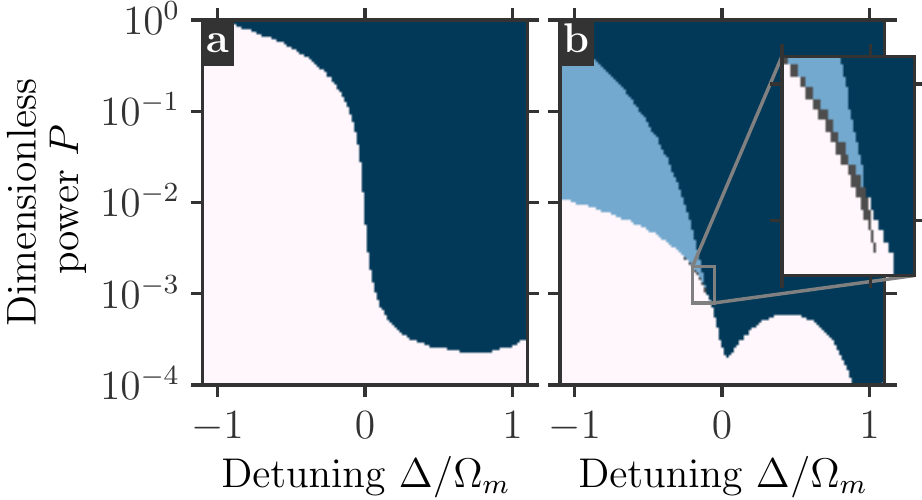}
	\endminipage
	\hfill{}
	\minipage{0.45\textwidth}
	\includegraphics[width=1\linewidth]{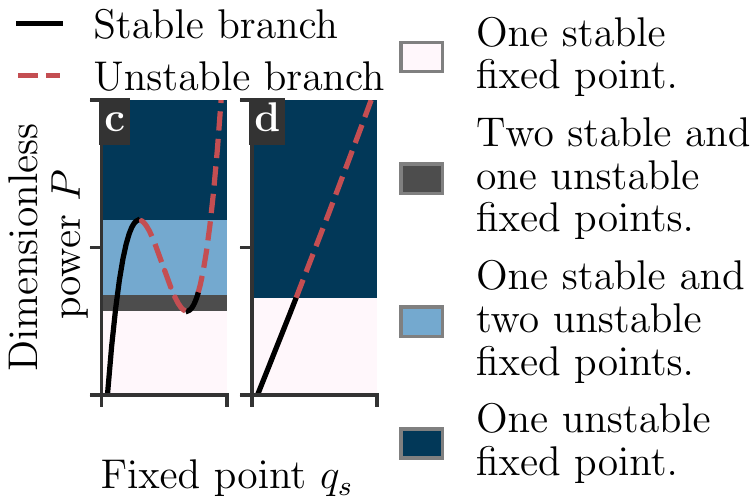}
	\endminipage 
	\caption{Stability diagram for an OM system. Panels (a) and (b) correspond
	to a strongly dissipative OM system with $\kappa=\Omega_{m}$ and
	$\gamma=10^{-3}\Omega_{m}$ and to a weakly dissipative one with $\kappa=0.1\Omega_{m}$
	and $\gamma=10^{-4}\Omega_{m}$, respectively. Colors mark different regions
	concerning the number and stability of the fixed points. Panels (c) and (d) illustrate
	the evolution of the fixed points with increasing $P$. Stable and
	unstable fixed points are represented by black and red dashed lines,
	respectively.}
	\label{stability_diag} 
\end{figure}

Although the fixed points of the OM systems have been known
for a long time~\cite{mancini1994,aspelmeyer2014},
it is important to understand them in more detail, because this will
provide the context for the discussions of the dynamical attractors.
The fixed point equations are obtained by setting the time derivatives
in Eqs.~(\ref{oc3},\ref{oc4}) to zero and solving the resulting
set of nonlinear equations: 
\begin{align}
	\alpha & = \left[2i\left(\frac{\Delta}{\Omega_{m}}+\sqrt{2}Q\right)-\frac{\kappa}{\Omega_{m}}\right]^{-1},
	\label{oc7}\\
	Q & =\frac{P}{\sqrt{2}}\left(1+\frac{\gamma^{2}}{4\Omega_{m}^{2}}\right)^{-1}|\alpha|^{2}.
	\label{oc8}
\end{align}
Here $Q=(\beta+\beta^{\ast})/\sqrt{2}$ is the rescaled position of
the mechanical oscillator. After inserting Eq.~(\ref{oc7}) into
(\ref{oc8}), we obtain a third order polynomial equation for $Q$
with real coefficients. Since $Q$ is also real, the system has at
least one fixed point; the maximum number is obviously three \cite{mancini1994}.
Figs.~\ref{stability_diag}(a,b) show the fixed point diagram for
an OM system with dissipation constants $\kappa=\Omega_{m}$ and $\gamma=10^{-3}\Omega_{m}$,
and for an OM system in the sideband-resolved regime, $\kappa=10^{-1}\Omega_{m}$
and $\gamma=10^{-4}\Omega_{m}$, respectively. Below, we will refer
to these two representative cases as the ``strongly dissipative''
and the ``weakly dissipative'' OM systems, respectively.
For sufficiently
small $P$ there is, as one would expect, just one stable fixed point.
As the parameter $P$ is increased, the fixed points can follow two
possible scenarios with different bifurcation phenomena. 
A bifurcation is a qualitative change of the dynamics which occurs as
a system parameter is varied~\cite{ott2002}. For fixed points, this typically means creation or
annihilation of fixed points, or change of the type of a fixed point (whether the 
fixed point is stable, unstable or hyperbolic).
The first scenario is shown in Fig.~\ref{stability_diag}
(c): an (inverse) saddle-node bifurcation%
\footnote{In a saddle-node bifurcation a pair of stable-unstable
fixed points approach each other as a parameter
$\eta$ is varied (for simplicity and without loss of generality, let us suppose that
we are increasing $\eta$). At $\eta = \eta^{\ast}$ the two fixed points merge and form one
single stable fixed point; if $\eta > \eta^{\ast}$ the fixed points cease to exist.
If $\eta$ is decreased one comes across the inverse saddle-node bifurcation, in which a pair
of stable-unstable fixed points is created.} 
takes place at some value of $P$ and
a pair of stable-unstable fixed points is created. Increasing
$P$ leads to a Hopf bifurcation%
\footnote{In the Hopf bifurcation a stable fixed point becomes unstable and a periodic
orbit appears as a parameter $\eta$ is varied. The periodic orbit can be unstable or stable.
In the latter case it is called a limit cycle.
The Hopf bifurcation is also known as a Poincar{\'e}-Andronov-Hopf bifurcation.}
at which the stable fixed point
becomes unstable. Further increase of $P$ results in a saddle-node
bifurcation at which a pair of stable-unstable fixed points is annihilated.
In some cases, the Hopf bifurcation may occur after the saddle-node
bifurcation. This scenario occurs only at $\Delta<0$ (``red detuning'').
The second scenario is shown in Fig.~\ref{stability_diag}(d): the
Hopf bifurcation again occurs at some value of $P$ and makes the
stable fixed point unstable. Further increase of $P$ does not change
the nature and the number of the fixed points.

Even though the above described bifurcations can be observed in both
strongly and weakly dissipative OM systems, Figs.~\ref{stability_diag}(a)
and \ref{stability_diag}(b) clearly show the essential difference
between them. When the dissipation is weaker, the bifurcations may
occur at much smaller values of $P$ and the stability diagram becomes
more complex. Since these bifurcations are genuine nonlinear phenomena
and $P$ is the strength of the nonlinear interaction, Figs.~\ref{stability_diag}(a,b)
provide us with a first indication that nonlinear effects are more
pronounced and even qualitatively altered in the weakly dissipative
case.

\subsection{Attractors}
The Liouville's theorem guarantees that the time evolution of Hamiltonian systems
preserves volumes in the phase space. In contrast to Hamiltonian systems,
dissipative systems are defined as systems in which volumes shrink over time
in some region of the phase space~\cite{ott2002}.
For these systems, generically speaking, the shrinking volumes collapse, in the long time limit,
to the so-called attractors. An attractor has the following properties~\cite{milnor1985}:\\
(i) It is a subset of the phase space which is invariant under the dynamics.\\
(ii) There must exist another (noninvariant) subset of the phase
space which defines the initial conditions for the trajectories asymptotically
approaching (being ``attracted'' by) the attractor at $t\to\infty$.
The second subset is called the basin of attraction.\\
(iii) An attractor cannot be decomposed in two or more disjoint
attractors.

The attractors of a dissipative system typically provide important
information about its dynamics. In particular, we expect them to illustrate the differences
between the strongly and weakly dissipative nonlinear dynamics of
OM systems. As said before, a stable fixed point is the simplest kind of attractor.
The Hopf bifurcation leads to the emergence of stable limit cycles,
which in turn can undergo transitions to other attractors, including
chaotic ones. In the strongly dissipative regime the limit cycles of a OM system
undergo the well known ``period-doubling cascade''%
\footnote{In a period-doubling bifurcation a stable
orbit with a period $T$ becomes unstable and a stable orbit with
period $2T$ appears as a parameter $\eta$ is varied.
A period-doubling cascade is an infinite sequence
of period-doubling bifurcations. The resulting stable
orbit does not have a finite period. Such orbits can be shown to
be chaotic attractors \cite{ott2002}.}
at $P\sim1$, becoming chaotic attractors. This phenomenon was described 
theoretically in Ref.~\cite{bakemeier2015} and 
observed in early pioneering experiments~\cite{carmon2007}.
In the weakly dissipative regime, however, where the fixed point analysis suggests
stronger nonlinear effects, neither the attractors nor the associated routes to chaos
have been studied. Below we focus on this regime.

\subsubsection{Basins of attraction and hypersensitivity to the initial state \label{HyperSens}}

A nonlinear dissipative system has generally more than one attractor
and its long time dynamics depends on initial conditions which can
belong to one or another basin of attraction. Some attractors can
be very challenging to reach both in numerical simulations and real
experiments because their basin of attraction is rather small and
their detection would require a nontrivial fine tuning of the initial
conditions. We will address the properties of those OM attractors
which are easily accessible and, therefore, relevant for experiments.
Throughout this section, we focus on the weakly dissipative case.

\begin{figure}[!ht]
	\centering
	\includegraphics[width=1\textwidth]{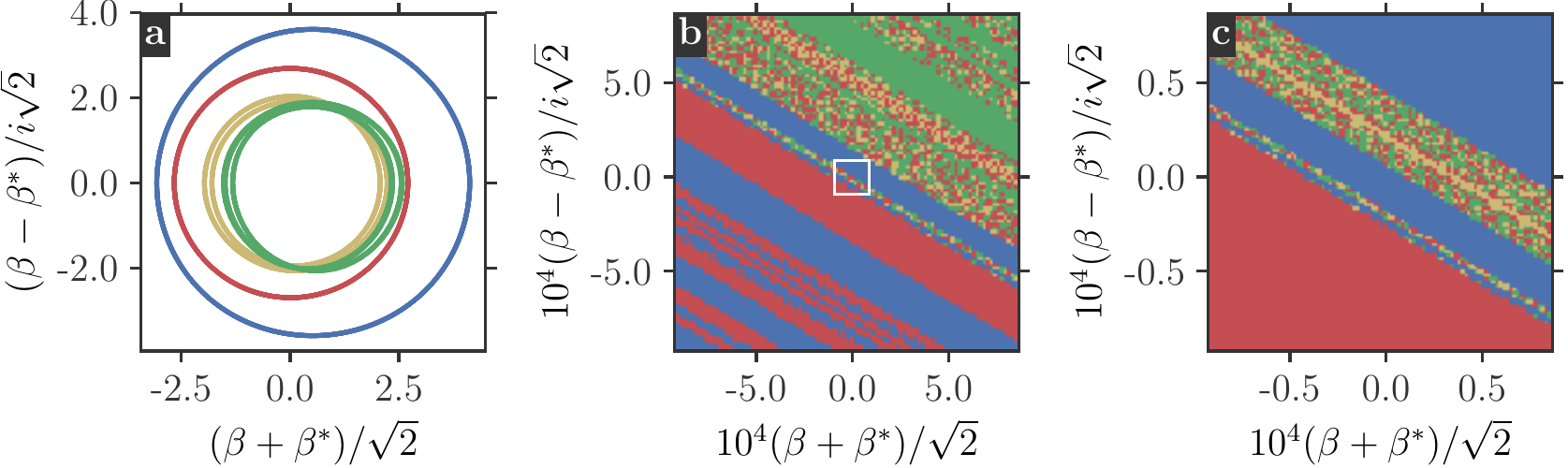}
	\caption{Panel~(a): Attractors of the OM system, projected into the mechanical
	phase space. We have chosen $\Delta=-0.754\Omega_{m}$ and $P=0.33$
	and detected limit cycles with periods 1 (blue and red lines), 2 (yellow
	line), and 4 (green line). Panel~(b): Basins of attraction close
	to the origin of the mechanical phase space. Colors correspond to
	the attractors shown in Panel~(a). Thermal mechanical fluctuations
	$\sigma_{\beta}=\sqrt{\left\langle \left|\beta\right|^{2}\right\rangle }=(g_{0}/\Omega_{m})\sqrt{n_{{\rm th}}}$
	would be on the order of $10^{-3}$ for realistic parameters with
	100 thermal phonons and $g_{0}/\Omega_{m}\sim10^{-4}$. Panel~(c):
	Zoom of the area within the white square shown in Panel~(b). The
	zoomed picture displays the same degree of complexity as in Panel~(b)
	and illustrate regions where the system is extremely sensitive even
	to minor changes in the initial conditions. The OM system operates in the
	weakly dissipative regime ($\kappa = 0.1 \Omega_m$ and 
	$\gamma = 10^{-4} \Omega_m$).}
	\label{attractors} 
\end{figure}

We have simulated Eqs.~(\ref{oc3},\ref{oc4}) for different initial
conditions of the mechanical oscillator%
\footnote{We have used the Julia package DifferentialEquations.jl \cite{rackauckas2010}
to obtain the numerical solution of the equations of motion. The numerical
integration method used is a 9th order Runge-Kutta method \cite{verner2010}
with relative tolerance set to $10^{-9}$ and absolute tolerance set
to $10^{-13}$.}, assuming that the laser is turned on abruptly at $t=0$ (thus $\alpha(0)=0$).
Fig.~\ref{attractors} shows the observed attractors and their basins
of attraction. While the strongly dissipative OM dynamics usually
reveals just one attractor, the phase space of the weakly dissipative
OM systems is much richer. One can observe not only several co-existing
attractors, i.e. multistability, but also very complex and entangled
basins of attraction, see Fig.~\ref{attractors}(b). Fig.~\ref{attractors}(c)
shows a zoom of a small part of the basin of attraction from Fig.~\ref{attractors}(b)
(the area within the white square) with a higher resolution. One
can see that, even on this scale, the basin of attraction is very
complex. This confirms that the weakly dissipative system possesses
hypersensitivity to the initial conditions. 

In a real experiment, the mechanical oscillator's initial state is
given by a thermal distribution at a given temperature ${\cal T}$.
In the classical regime studied here, one can use the Boltzmann (normal)
distribution with zero mean and variance
$\sigma_{m}^{2}=\left\langle \left|b\right|^{2}\right\rangle =k_{B}{\cal T}/\hbar\Omega_{m}$.
Note that, though the equations of motion (\ref{oc3},\ref{oc4})
contain only the parameter $P$, we will need also the OM coupling
$g_{0}$ to calculate the standard deviation of the dimensionless variable $\beta$:
$\sigma_{\beta}=\sqrt{\left\langle \left|\beta\right|^{2}\right\rangle }=(g_{0}/\Omega_{m})\sqrt{n_{{\rm th}}}$.
For a typical value $g_{0}=10^{-4}\Omega_{m}$ and a thermal phonon
number of 100, this amounts to $\sigma_{\beta}\sim10^{-3}$. As one
can see in the simulations, this standard deviation covers a range of different
attractors.

The hypersensitivity to the initial conditions hampers a comprehensive
study of the attractors of the weakly dissipative OM system. In addition
to analyzing dynamics for different values of $P$ and $\Delta$,
one would need also to consider many different initial conditions.
This can be computationally very expensive, especially in the presence
of chaotic attractors. The most common way to detect dynamical chaos
is to calculate the Lyapunov exponent (LE) of a given trajectory.
However, the convergence of the numerical methods available for calculating
the LEs is usually slow. This calls for the development of alternative
approaches. In the next section, we discuss such an alternative which
is faster, reliably detects the chaotic attractors, and moreover allows
one to determine the dimensionality of the regular attractors.

\section{The GALI method \label{GALI-sect}}

\subsection{Indicators of dynamical chaos}

An important task of any study of nonlinear dynamics is to distinguish
regular and chaotic parts of the phase space in the most efficient
way. A standard procedure for detecting chaotic trajectories is based
on calculations of the maximal Lyapunov exponent (mLE). Let us consider
the following general dynamical equations: 
\begin{equation}
	\frac{d}{dt}\vec{x}=\vec{F}(\vec{x}).
	\label{gali1}
\end{equation}
One can start from a given trajectory $\vec{x}(t)$ and focus on small
deviations $\vec{w}(t)$ from that trajectory. The linearized dynamics
of $\vec{w}(t)$ is described by 
\begin{equation}
	\frac{d}{dt}\vec{w}=J_{F}(\vec{x})\vec{w},
	\label{gali2}
\end{equation}
where $J_{F}(\vec{x})$ is the Jacobian matrix of $\vec{F}(\vec{x})$;
$\left[J_{F}\right]_{lj}=\partial F_{l}/\partial x_{j}$. The mLE $\lambda_{1}$
is defined as 
\begin{equation}
	\lambda_{1} = \lim_{t\to\infty}\Lambda(t),\
	\Lambda(t)=\frac{1}{t}\ln{\frac{|\vec{w}(t)|}{|\vec{w}(0)|}}.
	\label{gali3}
\end{equation}
Clearly, the mLE reflects the sensitivity of the trajectory $\vec{x}(t)$
to perturbations. A chaotic trajectory has positive mLE while regular
trajectories have nonpositive mLE, making $\lambda_{1}$ a good indicator
of chaotic dynamics. A numerical approximation for $\lambda_{1}$
can be obtained by calculating $\Lambda(t)$ in Eq.~(\ref{gali3})
for a sufficiently large $t$, at which $\Lambda(t)$ converges. This
approach, however, has the drawback that the convergence of $\Lambda(t)$
can be rather slow, and a long computation time is needed to learn
whether $\lambda_{1}$ is positive or not. Many chaos indicators have
been suggested to work around this problem; see Ref.~\cite{skokos2016}.
We have used two of them: the $SALI$ (Smaller ALignment Index) \cite{skokos2001}
and the $GALI$ (Generalized ALignment Index) \cite{skokos2007},
which are especially well-suited for our goals.

Before we discuss the $SALI$ and the $GALI$, we have to define all
LEs. Firstly, let us replace the n-dimensional vector $\vec{w}(t)$
in Eq.~(\ref{gali2}) by a $n\times n$ time-dependent matrix $W(t)$,
whose initial condition is $W(0)=\mathbb{1}$. The i-th column of
$W(t)$ describes the propagation of a perturbation acting in the
i-th direction of the phase space at $t=0$ (i.e. a perturbation proportional
to the vector with components $v_{j}=\delta_{j,i}$, where $\delta_{i,j}$
is the Kronecker delta). Using the singular value decomposition, one
can show that there is a set of $n$ nonnegative real numbers $\{\sigma_{1},\ldots,\sigma_{n}\}$,
and two sets of $n$ orthonormal vectors, $\{\vec{v}_{1},\ldots,\vec{v}_{n}\}$
and $\{\vec{u}_{1},\ldots,\vec{u}_{n}\}$, which satisfy the following
equation \cite{skokos2016}: 
\begin{equation}
	W(t)\vec{v}_{j} = \sigma_{j}\vec{u}_{j}.
	\label{gali4}
\end{equation}
This means that a perturbation in the direction of $\vec{v}_{i}$
at $t=0$ is mapped to a perturbation in the direction of $\vec{u}_{i}$
multiplied by $\sigma_{i}$ at time $t$. The definition of the LEs
reads 
\begin{equation}
	\lambda_{j} = \lim\limits _{t \to \infty}\frac{1}{t}\log{\sigma_{j}},
	\label{gali5}
\end{equation}
where $\{\sigma_{j}\}$ are sorted in decreasing order. Eq.~(\ref{gali5})
gives all LEs of the dynamical system, and not only $\lambda_{1}$.

Let us return to the n-dimensional vector $\vec{w}(t)$, which satisfies
Eq.~(\ref{gali2}). Using Eq. (\ref{gali4}), we can rewrite $\vec{w}(t)$
for $t\to\infty$ in the following way: 
\begin{equation}
	\vec{w}(t) = \sum_{j=1}^{n}(\vec{v}_{j},\vec{w}(0))\vec{u}_{j}e^{\lambda_{j}t},
	\label{gali6}
\end{equation}
where $\bigl(\vec{a},\vec{b}\bigr)$ denotes the inner product between
$\vec{a}$ and $\vec{b}$. Since $t$ is very large, the term proportional
to $e^{\lambda_{1}t}$ dominates the time dependence of $\vec{w}(t)$
(provided that $\lambda_{2}<\lambda_{1}$), such that Eqs.~(\ref{gali3})
and (\ref{gali5}) are consistent.

\begin{figure}
	\centering
	\minipage{0.495\textwidth} 
	\includegraphics[width=3in]{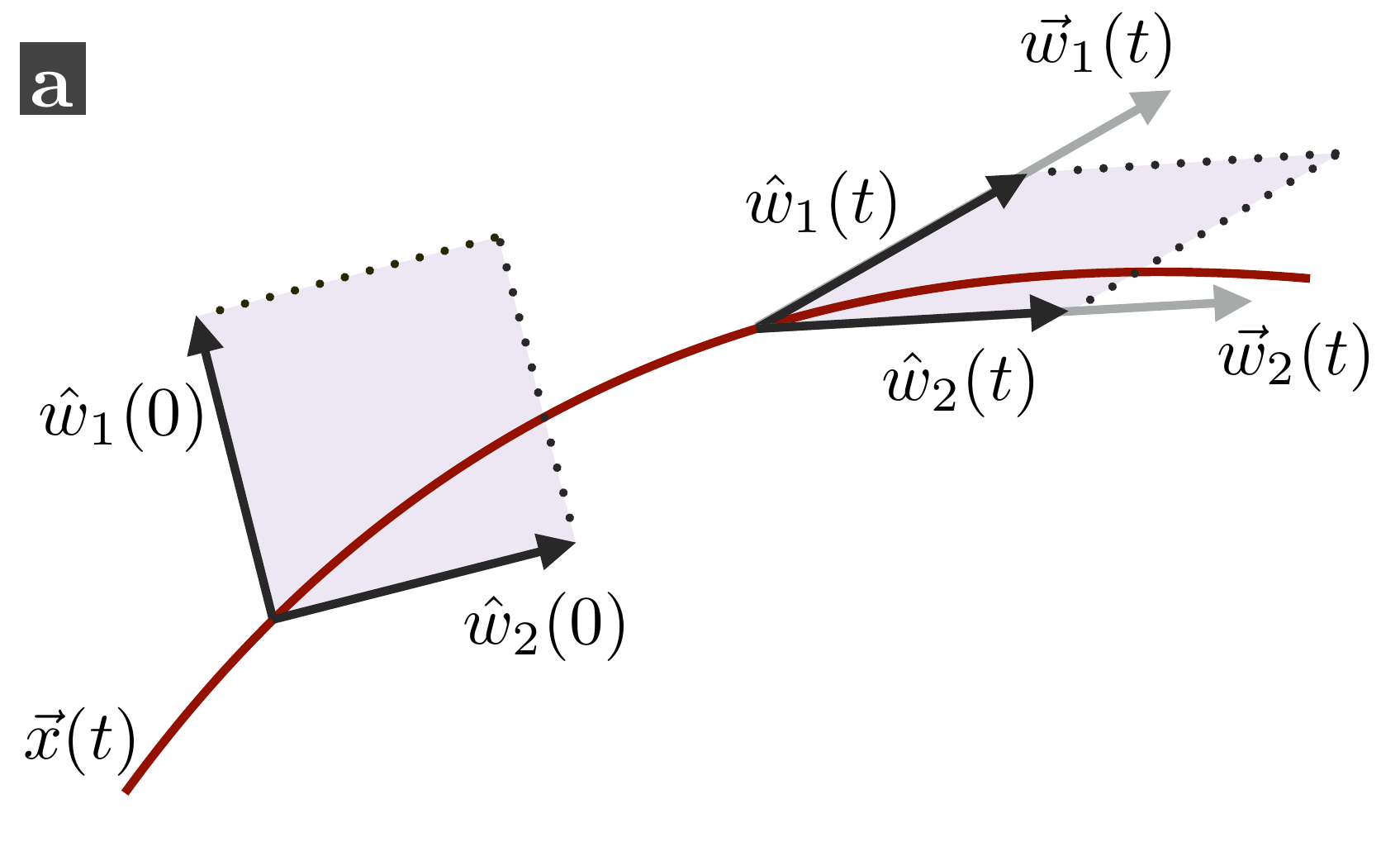}
	\endminipage
	\hfill{}
	\minipage{0.495\textwidth}
	\includegraphics[width=3in]{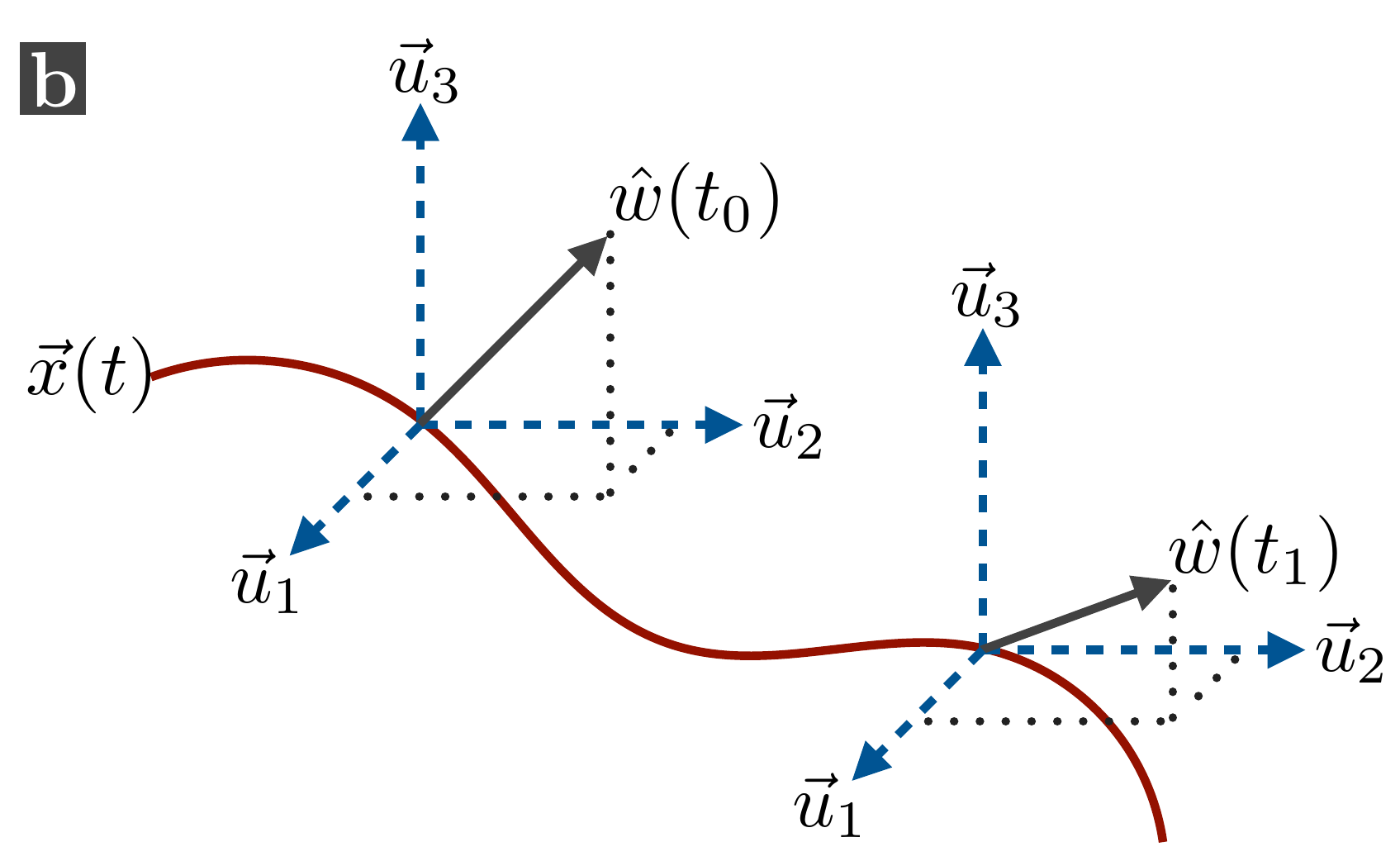}
	\endminipage 
	\caption{Panel~(a): Evolution of two deviation vectors along a chaotic trajectory.
	Even if $\vec{w}_{1}(0)\perp\vec{w}_{2}(0)$, the chaotic dynamics
	ensures that $\vec{w}_{1}(t)\parallel\vec{w}_{2}(t)$ at $t\to\infty$
	provided that $\lambda_{1}>\lambda_{2}$. Panel~(b): Evolution of
	the normalized deviation vector $\hat{w}(t)$ in the case $\lambda_{1}=\lambda_{2}$.
	When $t\to\infty$, $\hat{w}(t)$ approaches the plane defined by
	$\vec{u}_{1}$ and $\vec{u}_{2}$.}
	\label{gali_chaos} 
\end{figure}

Now, we are in a position to introduce the $SALI$ and the $GALI$.
These indicators of chaos have been initially suggested for Hamiltonian
systems, whose evolution preserves areas in the phase space. This
means that the LEs are either zero, or appear in pairs with the same
absolute value and opposite signs. The $SALI$ and the $GALI$ are
constructed in a similar way, but the $SALI$ is simpler; therefore,
we start with the $SALI$: Consider two orthogonal initial conditions
for Eq. (\ref{gali2}), $\vec{w}_{1}(0)\perp\vec{w}_{2}(0)$. Their
evolution yields vectors $\vec{w}_{1,2}(t)$ which become parallel
to $\vec{u}_{1}$, and consequently to each other, at $t\to\infty$; see
Fig.~\ref{gali_chaos}(a). This holds true if $\lambda_{1}>\lambda_{2}$ regardless
of the initial condition. The $SALI$ method uses this property to
distinguish the chaotic dynamics from the regular one. Let us define
\begin{equation}
	SALI(t)={\rm min}\{|\hat{w}_{1}(t)+\hat{w}_{2}(t)|,|\hat{w}_{1}(t)-\hat{w}_{2}(t)|\},
	\label{gali7}
\end{equation}
where $\hat{w}_{1,2}(t)=\vec{w}_{1,2}/|\vec{w}_{1,2}|$ are unit vectors,
and $(\vec{w}_{1}(0),\vec{w}_{2}(0))=0$. The above discussion suggests
that, if the dynamics is chaotic, the $SALI$ tends to zero as $t$
tends to infinity. In fact, the $SALI$ decays exponentially to zero
at the rate $\lambda_{1}-\lambda_{2}$ \cite{skokos2004}. If the
dynamics is regular, all LEs are zero, and there is no reason for
the alignment of vectors $\vec{w}_{1}(t)$ and $\vec{w}_{2}(t)$.
The $SALI$ does not decay to zero in this case. 

Thus, the $SALI$ is a good chaos indicator for Hamiltonian systems
provided that $\lambda_{1}\neq\lambda_{2}$. In the opposite case,
where $\lambda_{1}=\lambda_{2}$, Eq.~(\ref{gali6}) suggests that
$\vec{w}(t)$ tends to $c_{1}\vec{u}_{1}+c_{2}\vec{u}_{2}$, with
$c_{1,2}$ depending on $\vec{w}(0)$. Therefore, $\vec{w}_{1}(t)$
and $\vec{w}_{2}(t)$ do not become parallel at $t\to\infty$ but
rather approach the plane defined by $\vec{u}_{1}$ and $\vec{u}_{2}$;
see Fig.~\ref{gali_chaos}(b). The $SALI$ does not decay to zero
and a more advanced chaos indicator is needed. To construct it, we
calculate the time evolution of a third deviation vector, $\vec{w}_{3}(t)$,
satisfying $(\vec{w}_{3}(0),\vec{w}_{1,2}(0))=0$. We then compute
the volume of the parallelepiped defined by the vectors $\hat{w}_{1,2,3}(t)$.
It is given by the so-called $GALI_{3}$: 
\begin{equation}
	GALI_{3}(t) = |\hat{w}_{1}(t)\wedge\hat{w}_{2}(t)\wedge\hat{w}_{3}(t)|.
	\label{gali8}
\end{equation}
Here $\hat{w}_{i}=\vec{w}_{i}/|\vec{w}_{i}|$ is again the unit vector,
and $\vec{a}\wedge\vec{b}$ is the exterior product between the vectors
$\vec{a}$ and $\vec{b}$. One can show that $GALI_{3}\propto\exp(-2\lambda_{1}t+\lambda_{2}t+\lambda_{3}t)$
\cite{skokos2007}, and it decays to zero exponentially quickly unless
$\lambda_{1}=\lambda_{2}=\lambda_{3}$. It can be shown that the $GALI_{3}$
decays to zero also on some regular orbits. However, such a non-chaotic
decay is much slower as it follows a power law. This allows one to
distinguish the chaotic and regular motion \cite{skokos2007}.

If the first $(k-1)$ LEs are equal to each other and positive, the
chaotic and regular motion are distinguished by the $GALI_{k}$ \cite{skokos2007}:
\begin{equation}
	GALI_{k}(t)=|\hat{w}_{1}(t)\wedge\ldots\wedge\hat{w}_{k}(t)|
	\propto \exp[-(\lambda_{1}-\lambda_{2})t-(\lambda_{1}-\lambda_{3})t-\ldots-(\lambda_{1}-\lambda_{k})t].
	\label{gali81}
\end{equation}
It is clear that $GALI_{k}(t)$ does not decay exponentially if and
only if $\lambda_{1}=\lambda_{2}=\ldots=\lambda_{k}$. This applies
to regular orbits where $\lambda_{1\ldots k}=0$. If the trajectory
is chaotic, there exists a $k$ which is smaller than the phase space
dimension such that $GALI_{k}$ decays exponentially. One can show
that $SALI\propto GALI_{2}$ \cite{skokos2007}. Therefore, we will
refer only to the $GALI$ in what follows.

\subsection{The $GALI$ method for dissipative systems}

We have already mentioned that the $GALI$ has been developed as an
indicator of chaos for Hamiltonian systems, and its archetypal treatment
generally does not work in the presence of dissipation and attractors.

Before extending the $GALI$ to dissipative dynamics, let us first
comment on the relation between attractors and LEs.
The ``attraction'' of nearby orbits by the attractor comes from
the fact that some LEs are negative (when the system is near the attractor).
If the attractor is regular, all
LEs are non-positive, and the number of zero-valued LEs is equal to
the dimension of the attractor, see Chapter 10 of Ref.~\cite{anishchenko2014}.
If all LEs are negative, the attractor is a fixed point. An attractor
with only one zero-valued LE is a 1D curve in phase space, that is
commonly called a limit cycle. An attractor which has $p$ zero-valued
LEs is a $p$-dimensional torus in phase space, that is dubbed a limit
torus. The most complex attractors have positive and
negative LEs, such that ``attraction'' co-exists with chaotic divergence
of the trajectories. Those are called chaotic or strange attractors.

Consider now the $GALI_{2}$ in a dissipative system. On a limit cycle,
$\vec{w}(t)$ approaches $\vec{v}_{1}$ (regardless of the initial
condition), and the $GALI_{2}$ decays exponentially to zero at a
rate $-\lambda_{2}$. On the other hand $\vec{w}(t)$ approaches
$\vec{v}_{1}$ on a chaotic attractor with $\lambda_{1}>\lambda_{2}$
(again regardless of the initial conditions) and $GALI_{2}$ decays
exponentially to zero at a rate $\lambda_{1}-\lambda_{2}$. We can
conclude that the $GALI_{k}$ decays to zero both on the limit cycle
and on the chaotic attractor for all possible values of $k$. Therefore,
the $GALI$ method cannot distinguish between the limit cycle and
the chaotic attractor. We argue that the $GALI$ is nevertheless useful
for the study of dissipative systems because it is able to distinguish
dynamics in the vicinity of the attractor from transient dynamics.
Let us use Eqs.~(\ref{gali6},\ref{gali81}) to analyse the behaviour
of the deviation vector modulus, $|\vec{w}(t)|$, and of the $GALI_{k}$
on the different kinds of attractors: 
\begin{itemize}
	\item \textit{Fixed point}: all LEs are negative. Consequently, $|\vec{w}(t)|$
	decays to zero exponentially quickly. The $GALI_{k}$ do not necessarily
	decay to zero since some of the LEs may have the same value. 
	\item \textit{Limit cycle}: $\lambda_{1}=0$, while all other LEs are negative.
	Consequently, $|\vec{w}(t)|$ does not decay to zero. The $GALI_{k}$,
	on the other hand, decay to zero exponentially quickly for $k\ge2$. 
	\item \textit{p-dimensional limit torus}: $\lambda_{1}=\ldots=\lambda_{p}=0$,
	while all other LEs are negative. Consequently, $|\vec{w}(t)|$ and
	the $GALI_{k}$ for $k\le p$ do not decay to zero. The $GALI_{k}$
	for $k>p$, on the other hand, decays to zero exponentially quickly. 
	\item \textit{Chaotic attractor}: there are generically $N_{1}$ positive
	LEs and $N_{2}$ negative ones, where $N_{1,2}>0$. Consequently,
	$|\vec{w}(t)|$ grows and the $GALI_{k>N_{1}}$ decay exponentially
	quickly. The behaviour of the $GALI_{2\le k\le N_{1}}$ (whether or
	not they decay to zero) depends on the degeneracy of the positive LEs.
\end{itemize}
Hence, when the trajectory is in the vicinity of an attractor, either
$|\vec{w}(t)|$ or some $GALI_{k}$ must decay exponentially. Note
that the inverse statement does not hold true: the fast decay of either
$|\vec{w}(t)|$ or some $GALI_{k}$ cannot prove that the trajectory
is in the vicinity of the attractor.

The transient dynamics is more difficult for the analysis since one
cannot make any general statement about the behaviour of $|\vec{w}(t)|$
or the $GALI_{k}$ when the trajectory is not close to any attractor.
In principle, there is a possibility that $|\vec{w}(t)|$ or the $GALI_{k}$
could decay to very small values during the transient dynamics. On
the other hand there is no generic reason for such a behaviour and
it seems unlikely that many different deviation vectors would behave
in such a way. Therefore, we will assume that whenever either $|\vec{w}(t)|$
or the $GALI_{k}$ decays to zero, the trajectory is in the vicinity
of an attractor.

Once we know that the trajectory is in the vicinity of the attractor,
knowing the properties of $|\vec{w}(t)|$ suffices to distinguish
the chaotic attractors from the regular ones. If $|\vec{w}(t)|$
grows exponentially the attractor is chaotic; if there is no
exponential growth of $|\vec{w}(t)|$ the attractor is regular. In
the latter case, the $GALI_{k}$ provides the information about the
dimensionality of the attractor. The ability of the $GALI$ to detect
the transient dynamics is especially important for a blue detuned
OM system, since deterministic (non-chaotic) amplification of the
mechanical motion represents the default behaviour in this regime
and the growth of $|\vec{w}(t)|$ could be easily misinterpreted as
a signature of chaos.

Armed with this novel understanding, we have successfully applied
the $GALI$ method to the dynamics of weakly dissipative OM systems.
This will be the focus of the next section.

\section{Applying the $GALI$ method to OM systems \label{GALI-OM}}

\subsection{Details of the implementation}

In the previous Section, we have explained that the $GALI$ method
is a powerful tool for the analysis of the attractors of weakly dissipative
OM systems because it allows one to detect the chaotic attractors
very efficiently and to distinguish the regular attractors of different
dimensionality. To study the nonlinear OM dynamics, we have solved
the equations of motion (\ref{oc3},\ref{oc4}) and analyzed the evolution
of three deviation vectors $\vec{w}_{1,2,3}(t)$ whose initial conditions
are orthogonal. After this, we have calculated the $GALI_{2,3}(t)$.
Three different pairs chosen from the three deviation vectors can
generate three $GALI_{2}$. We have calculated the $GALI_{2}^{(w_{1},w_{2})}(t)$
based on $\vec{w}_{1,2}(t)$ and the $GALI_{2}^{(w_{1},w_{3})}(t)$
based on $\vec{w}_{1,3}(t)$. We have used the average norm, 
\begin{equation}
	\langle w(t)\rangle = \left[ |\vec{w}_{1}(t)|+|\vec{w}_{2}(t)|+|\vec{w}_{3}(t)| \right]/3,
\end{equation}
the average $GALI_{2}$, 
\begin{equation}
	\langle GALI_{2}(t)\rangle = \frac{1}{2}\left[ GALI_{2}^{(w_{1},w_{2})}(t)+GALI_{2}^{(w_{1},w_{3})}(t) \right],
\end{equation}
and the $GALI_{3}(t)$ for classification of the attractors. Specifically,
we have assumed that any of these quantities has effectively ``decayed
to zero'' when it becomes smaller than a given cutoff $\epsilon$.
We have chosen $\epsilon=10^{-6}$. Our operational rules are: 
\begin{itemize}
	\item If $\langle w(t)\rangle<\epsilon$, the attractor is a \emph{fixed point}. 
	\item If $\epsilon\ll\langle w(t)\rangle\ll\epsilon^{-1}$ and $\langle GALI_{2}(t)\rangle<\epsilon$,
	the attractor is a \emph{limit cycle}. 
	\item If $\epsilon\ll\langle w(t)\rangle,\langle GALI_{2}(t)\rangle\ll\epsilon^{-1}$
	and $GALI_{3}(t)<\epsilon$, the attractor is a \emph{2-dimensional limit torus}%
	\footnote{When the attractor is a limit cycle, the $GALI_{3}$ frequently decays
	much faster than the $GALI_{2}$. For this reason, one can erroneously
	conclude that the attractor is a 2-dimensional torus. To avoid this
	mistake, one should calculate the $GALI_{2}$ for a longer time. This
	will reliably detect the cases where the attractor is a limit cycle,
	and not a torus.}. 
	\item If $\langle w(t)\rangle>\epsilon^{-1}$ and either $\langle GALI_{2}(t)\rangle<\epsilon$
	or $GALI_{3}(t)<\epsilon$, the attractor is \emph{chaotic}. 
\end{itemize}
Note that, since the OM phase space is 4-dimensional, we could, in
principle, come across limit tori with higher dimensionality. Their
detection would require using the fourth vector $\vec{w}_{4}(t)$
and constructing $GALI_{4}(t)$ because neither $\langle w(t)\rangle$
nor $\langle GALI_{2}(t)\rangle$ nor $GALI_{3}(t)$ would drop below
$\epsilon$. We will show, however, that this is not the case for
our choice of the parameters and of the initial conditions and, thus,
the selected indicators suffice for our purposes.

\subsection{Attractors of weakly dissipative OM systems}

\begin{figure}[!ht]
	\includegraphics[width=1\linewidth]{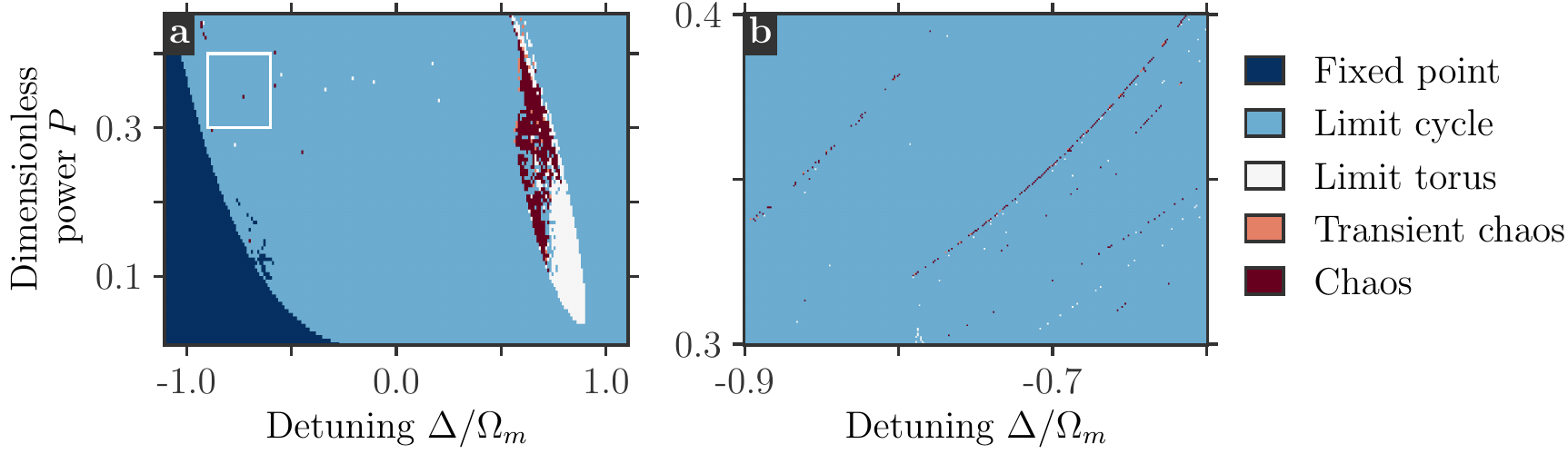} 
	\centering
	\includegraphics[height=2in]{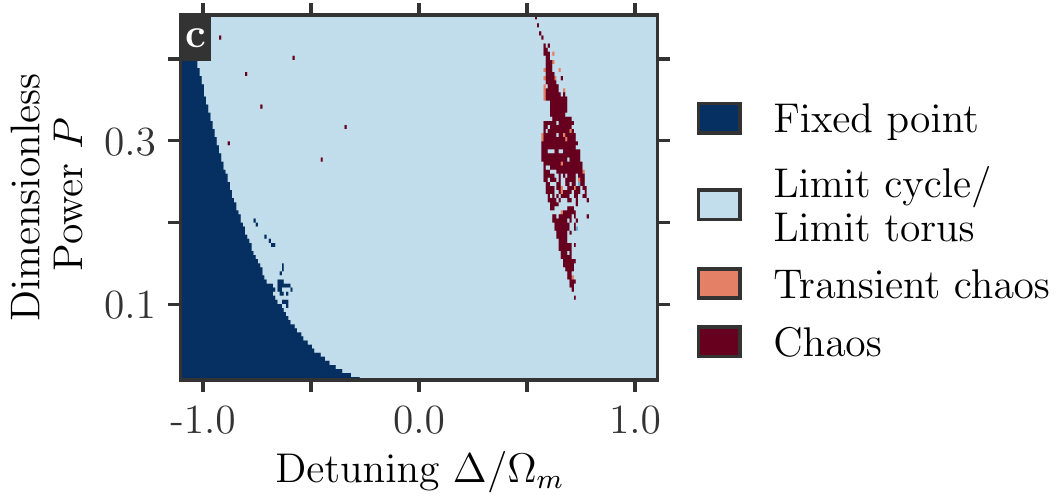}
	\caption{Attractor diagrams for the weakly dissipative OM system which have
	been generated by using the $GALI$ method (Panels (a) and (b)) and
	the mLE method (c). Each pixel has been obtained after solving the
	equations of motion for ten different initial conditions. Its color
	corresponds to the the most ``complex'' attractor which we have
	detected for given $\Delta$ and $P$. Note that the mLE method does
	not distinguish between the limit cycles and the limit tori. Panel~(b)
	shows a zoom of the area within the white box in Panel~(a). The OM 
	system operates in the weakly dissipative regime ($\kappa = 0.1 \Omega_m$
	and $\gamma = 10^{-4} \Omega_m$).}
	\label{chaos_diagram} 
\end{figure}

Fig.~\ref{chaos_diagram}(a) shows a diagram as a function of the detuning
$\Delta$ and the drive power $P$ which confirms the existence of various
attractors in the phase space of a weakly dissipative OM system. We
have already discussed that OM systems possess multistability: several
attractors of different dimension can co-exists at given values of
$\Delta$ and $P$. Therefore, each pixel of the diagram has been
obtained by solving the equations of motion for ten different initial
conditions. Its color corresponds to the most ``complex'' attractor
observed in these ten simulations. The attractors, sorted by increasing
``complexity'', are: fixed points, limit cycles, limit tori, transiently
chaotic attractors, and chaotic attractors. 

To prove that our implementation of the $GALI$ method yields reliable
results, we show in Fig.~ \ref{chaos_diagram}(c) a similar diagram
which has been obtained by calculating the mLE. One can observe qualitative
similarity of the results generated by the two different methods,
which confirms the validity of the diagram \ref{chaos_diagram}(a).
On the other hand, this comparison also shows that the mLE method
yields less detailed information and is unable to distinguish between
the limit cycles and the limit tori. 

\begin{figure}[!ht]
	\minipage{0.5\textwidth}
	\includegraphics[width=1\linewidth]{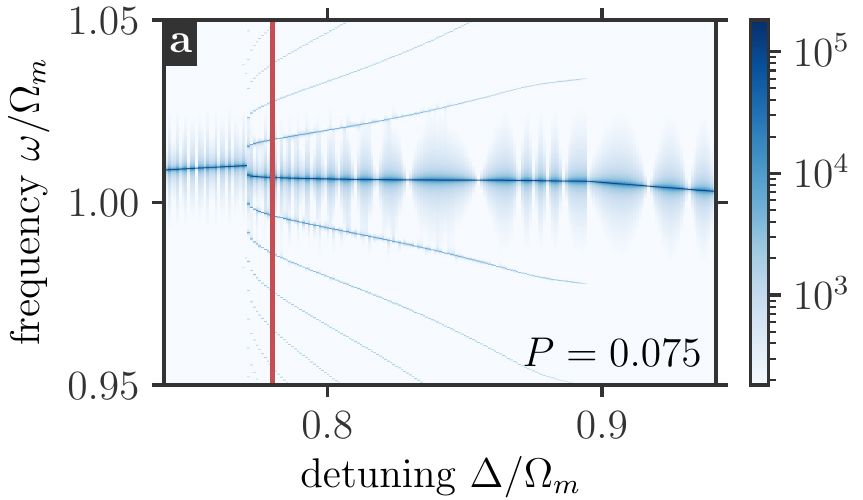}
	\endminipage
	\hfill{}
	\minipage{0.5\textwidth}
	\includegraphics[width=1\linewidth]{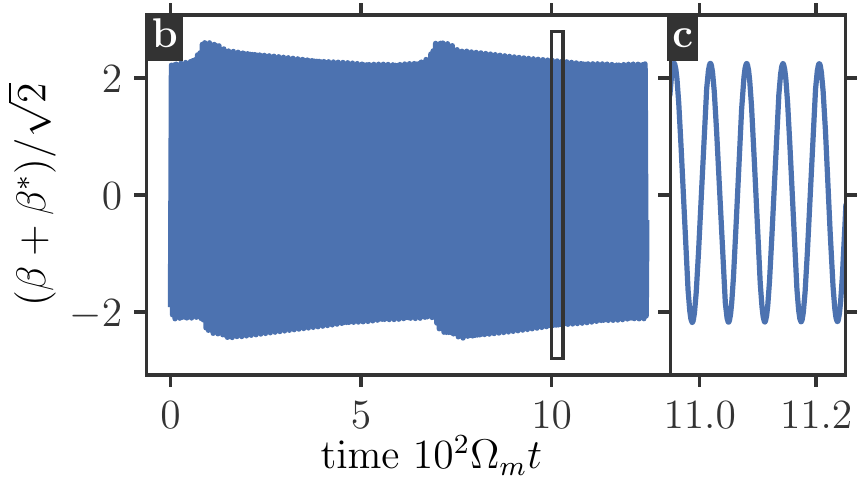}
	\endminipage 
	\caption{Panel~(a): The mechanical spectrum for an OM system close to an attractor. We
	have initially simulated the dynamics for $P=0.075\Omega_{m}$ and
	$\Delta=0.74\Omega_{m}$ until the system reached an attractor. Afterwards,
	the detuning $\Delta$ has been slowly increased while $P$ was kept
	fixed. The spectrum of the position of the mechanical oscillator has
	been computed during this process. Colors denote the absolute value
	of the spectrum, $|S(\omega)|$. Only one peak is observed at $\Delta<0.77\Omega_{m}$,
	i.e. the attractor is a limit cycle. Several other peaks appear at
	$\Delta \simeq 0.77\Omega_{m}$, i.e. the attractor becomes a limit
	torus. These secondary peaks disappear at $\Delta \simeq 0.89\Omega_{m}$;
	for $\Delta>0.89\Omega_{m}$, the attractor is again a limit cycle.
	Panel~(b): Time evolution of the mechanical degree of freedom;
	the system is in the vicinity of the limit torus marked
	by the red line in Panel~(a). The beating created by the sidebands is visible.
	Panel~(c): Zoom of the area within the black box in Panel~(b).}
	\label{NSbifurcation} 
\end{figure}

\subsection{Limit tori, Neimark-Sacker bifurcation, and transition to chaos}

The diagram \ref{chaos_diagram}(a) displays the presence of four
OM attractors with different dimensions: fixed points, limit cycles,
limit tori and chaotic attractors. While there is a number of works
addressing OM limit cycles (see e.g. Refs.~\cite{marquardt2006, ludwig2008,
loerch2014, wurl2016, kippenberg2005, metzger2008, weiss2016}),
and some works devoted to chaos in OM systems (see e.g. Refs.~\cite{bakemeier2015,
djorwe2018, carmon2007, monifi2016, mwang2016, wu2017, 
navarro-urrios2017, jin2017, wang2016, lue2015}),
studies of the OM limit tori, or quasiperiodic orbits, are scarce.
We are aware of only one paper, Ref.~\cite{wang2014}, which reports
the theoretical prediction of quasiperiodic OM orbits for parameters close to
our choice. Quasiperiodic orbits were not observed in the strongly
dissipative OM system, cf. Ref.~\cite{bakemeier2015}.

We have detected the limit tori mostly in the range $0.6\Omega_{m}\le\Delta\le\Omega_{m}$,
which corresponds to the blue-detuned regime. The limit tori can be
also found in the red detuned region, but they are rather rare there.
Fig.~\ref{NSbifurcation}(a) shows how a quasiperiodic orbit appears
and disappears when the detuning is changed adiabatically%
\footnote{Adiabatic change here means that the detuning was changed very slowly,
such that if the system is initially close to some attractor, it remains close to it.}.
We have plotted the spectrum of the position of the mechanical oscillator
when the OM system is close to some attractor. There is only one peak
in the spectrum at $\Delta\sim0.7\Omega_{m}$ which means that the
attractor is a limit cycle. A qualitative change occurs at $\Delta\simeq0.77\Omega_{m}$
and several peaks appear at larger $\Delta$. The motion is quasiperiodic
in this range and the attractor is now a limit torus%
\footnote{The presence of the secondary peaks does not necessarily imply that
the attractor is a torus. For this to happen, two frequencies in the
spectrum must be incommensurate. We have concluded that the attractor
is indeed the limit torus because the $GALI_{2}$ does not decay to
zero.}.
All secondary peaks disappear at $\Delta\simeq0.89\Omega_{m}$,
and again only one peak is visible%
\footnote{The low-intensity semi-periodic pattern around
the main peak is a numerical artefact connected to the way the Fourier
transform was implemented.};
the attractor becomes a limit cycle at $\Delta>0.89\Omega_{m}$.
These two transitions between a limit cycle
and a limit torus agree with the diagram \ref{chaos_diagram}(a) and are known in the
literature as the Neimark-Sacker bifurcation \cite{neimark1959,sacker1964}.
Fig.~\ref{NSbifurcation}(b) shows time evolution of the mechanical
degree of freedom. This trajectory is in the vicinity of a limit torus. The
beating created by the sidebands is clearly visible.
A similar phenomenon has been observed in Refs.~\cite{bagheri2013, seitner2017}.

Our remarkable finding is that the critical value of $P$, at which
chaos appears, becomes considerably smaller when the dissipation is
weak; compare the value $P_{c}\approx0.1$ from Fig.~\ref{chaos_diagram}(a)
with $P_{c}\approx1.4$ reported in Ref.~\cite{bakemeier2015} for
the strongly dissipative case. We have
discovered another qualitative difference between the strongly and
weakly dissipative chaotic OM dynamics: chaos is observed mostly in
the red detuned regime ($\Delta<0$) in the former case, while in
the latter case it is observed mostly in the blue detuned regime ($\Delta>0$).
In order to support the claim that the differences between the 
strongly and the weakly dissipative regimes depend on
the sideband parameter $\kappa/\Omega_m$ only, we have obtained
the attractor diagram also for $\kappa = 10^{-1} \Omega_m$ and
$\gamma = 10^{-3} \Omega_m$. This diagram, which is not shown here, displays
the same qualitative features observed in Fig.~\ref{chaos_diagram}(a),
differing markedly from the results
reported in \cite{bakemeier2015}. This shows that the sideband parameter
$\kappa/\Omega_m$ is the only relevant parameter in our classification of
strongly and weakly dissipative regimes.


Though we have detected chaos for both positive and negative detuning,
the chaotic region in the blue detuned part of the diagram looks more
``dense'' because pixels denoting chaotic dynamics agglomerate and
are not isolated. We expect that small changes of the parameters inside
the agglomerates cannot destroy chaotic dynamics. The chaotic region
in the red detuned regime is very sparse and even subtle changes of
the parameters are likely to convert the chaotic attractor to a regular
one. Such a ``sparse chaotic region'' is shown in Fig.~\ref{chaos_diagram}(b)
which displays a zoomed part of Fig.~\ref{chaos_diagram}(a) (the
area within the white box in the red detuned region). One can see that the
sparse chaotic region consists of very thin chaotic layers.

\begin{figure}[!ht]
	\centering
	\includegraphics[width=1\textwidth]{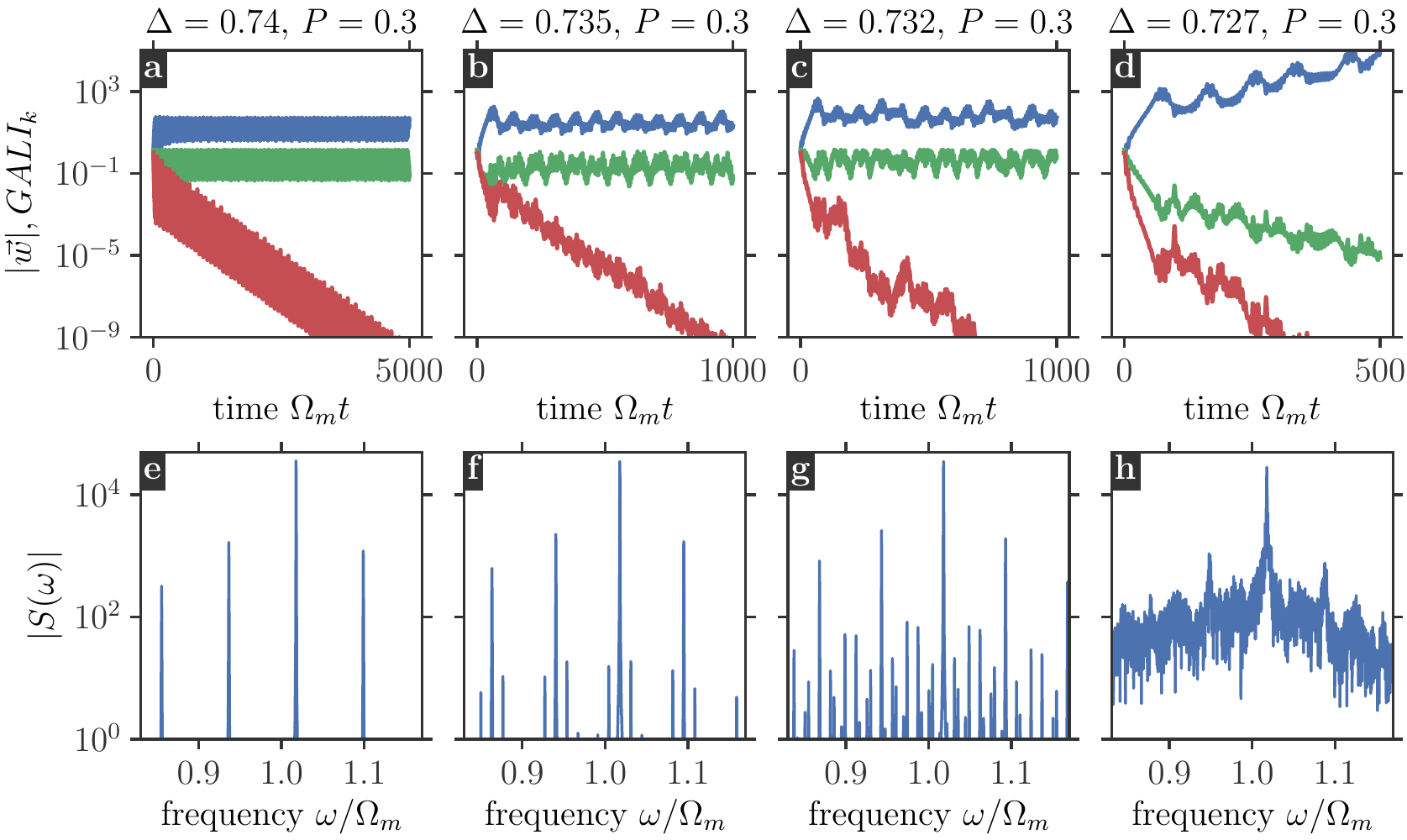}
	\caption{Transition from a limit torus to a chaotic attractor when the detuning
	$\Delta$ is changed adiabatically. Panels~(a-d) show the time evolution
	of the modulus of the deviation vector, $|\vec{w}|$, (blue curve),
	the $GALI_{2}$ (green curve), and the $GALI_{3}$ (red curve). Panels~(e-h)
	display the spectrum of the position of the mechanical oscillator.
	The power is kept fixed at $P=0.3$ for all figures while the detuning
	$\Delta$ is changed. Panels~(a,e), $\Delta=0.74\Omega_{m}$: The
	attractor is a 2-dimensional torus because only $GALI_{3}$ decays
	to zero exponentially quickly. The spectrum shows only two frequencies.
	Panels~(b,f), $\Delta=0.735\Omega_{m}$, and (c,g), $\Delta=0.732\Omega_{m}$:
	The attractor remains a 2- dimensional torus though the spectrum contains
	more frequencies with decreasing $\Delta$. Panels~(d,h), $\Delta=0.727\Omega_{m}$:
	The spectrum is dense; the $GALI_{2,3}$ decay to zero while $|\vec{w}|$
	increases exponentially quickly. Hence, we have come across a chaotic attractor.}
	\label{quasiperiodic} 
\end{figure}

Close proximity of chaotic and quasiperiodic regions in the diagram
Fig.~\ref{chaos_diagram}(a) at $\Delta>0$ provides a hint that
OM systems can reach dynamical chaos via a route involving quasiperiodic
orbits. To test this guess, we have investigated how an OM attractor
behaves when the detuning is changed adiabatically such that the system
starts in a quasiperiodic region of the parameters space and ends
in a chaotic region. The results are shown in Fig.~\ref{quasiperiodic}.
The time evolution of $|\vec{w}|$, the $GALI_{2}$, and the $GALI_{3}$
are given in the upper Panels, while the lower Panels present the
spectrum of the motion of the mechanical oscillator. At $\Delta=0.74\Omega_{m}$,
in Panels~(a,e), $|\vec{w}|$ and $GALI_{2}$ oscillate around some
nonzero values, while $GALI_{3}$ decays to zero exponentially quickly.
Simultaneously, the mechanical spectrum has only two independent frequencies.
Therefore, the attractor is a 2-dimensional torus. When $\Delta$
is decreased (down to $\Delta=0.735\Omega_{m}$, Panels~(b,f), and
further to $\Delta=0.732\Omega_{m}$, Panels~(c,g)), the behaviour
of all three indicators remains qualitatively the same though more
and more additional peaks (marking more frequencies) become visible
and pronounced in the spectrum. The dynamical picture becomes qualitatively
different at the smallest chosen detuning ($\Delta=0.727\Omega_{m}$,
Panels~(d,h)): $|\vec{w}|$ increases while the $GALI_{2,3}$ decay
to zero exponentially. It means that the attractor is chaotic. This
conclusion is confirmed by the dense nature of the mechanical spectrum.
The transition to chaos depicted in Fig.~\ref{quasiperiodic} is
called \textit{the quasiperiodic route to chaos} \cite{dixon1996}.
It is characterized by the appearance of new frequencies when the
control parameter ($\Delta$ in our study) is changed. The new frequencies
must be commensurate with the basic two frequencies, see Fig.~\ref{quasiperiodic}(e,f,g).
If the new frequencies were incommensurate we would come across a
higher dimensional torus and the $GALI_{3}$ would not vanish. One
can notice a similarity between the quasiperiodic route to chaos and
the period doubling cascade. Indeed, the dense chaotic spectrum is
reached via an increasing number of new frequencies which are commensurate.

\subsection{Classical transient chaos}

Let us finally discuss another nonlinear phenomenon, which is captured
by Fig.~\ref{chaos_diagram}(a) but has not been revealed in previous
studies of classical OM chaos. This is the well-known transient chaos.
Its name perfectly reflects its main features: a dynamical system
can display chaotic motion for a finite time interval after which
its dynamics becomes regular. Dissipative transient chaos can be explained
by a coexistence of different attractors, e.g. one attractor is chaotic
and the other regular. Each attractor has its own basin of attraction.
The basins could be separated by only an unstable periodic orbit.
One can tune a parameter of the nonlinear system, $\eta$, such that
the chaotic attractor approaches the unstable periodic orbit. At a
critical value $\eta=\eta_{c}$, the chaotic attractor ``touches''
the unstable periodic orbit. This phenomenon is called the boundary
crisis \cite{grebogi1982,grebogi1983} and it is one possible mechanism
underlying transient chaos. One can imagine that, at $\eta \gtrsim \eta_{c}$,
a tiny fraction of the chaotic attractor penetrates the basin of attraction
of the regular attractor. If a chaotic trajectory reaches this intersection
region, where the chaotic attractor is entangled with the regular
basin of attraction, it can be intercepted and ``dragged'' into
the regular attractor. In other words, the chaotic attractor becomes
leaky%
\footnote{This discussion is, of course, not rigorous, rather illustrative.
From a mathematical point of view, the chaotic attractor ceases to
exist at $\eta>\eta_{c}$.}.

Transient chaos also provides a possible route to chaos: changing
$\eta$ in the opposite direction results in the creation of a chaotic
attractor at $\eta \lesssim \eta_{c}$. The time that a trajectory spends
on the chaotic attractor before the leakage is typically very sensitive
to the initial conditions. Nevertheless, we can define the average
escape time $\tau_{{\rm esc}}$. To this end, we select $N_{0}$ points
in the leaky attractor, and use them as initial conditions of the
nonlinear system. We then compute $N(t)$, the number of trajectories
remaining on the chaotic attractor at time $t$. The escape time can
be found from the approximation $N(t)\simeq N_{0}\exp(-t/\tau_{{\rm esc}})$.
Clearly, numerical approaches cannot distinguish the genuine chaotic
trajectories and the transient trajectories with very large $t_{{\rm esc}}$%
\footnote{We distinguish here the average escape time $\tau_{{\rm esc}}$, which
is a property of the chaotic attractor, and the escape time $t_{{\rm esc}}$,
which is a property of the particular trajectory.}.
We have used an empirical criterion: (i) trajectories which display
chaotic motion during a time interval larger than $T_{cutoff}=10^{5}\Omega_{m}^{-1}$
are labeled ``chaotic''; (ii) trajectories whose dynamics remains
chaotic only for shorter times and becomes regular afterwards are
labeled ``transiently chaotic''. Interested readers can find more
details on transient chaos in the book \cite{lai2011}.

\begin{figure}
	\includegraphics[width=1\linewidth]{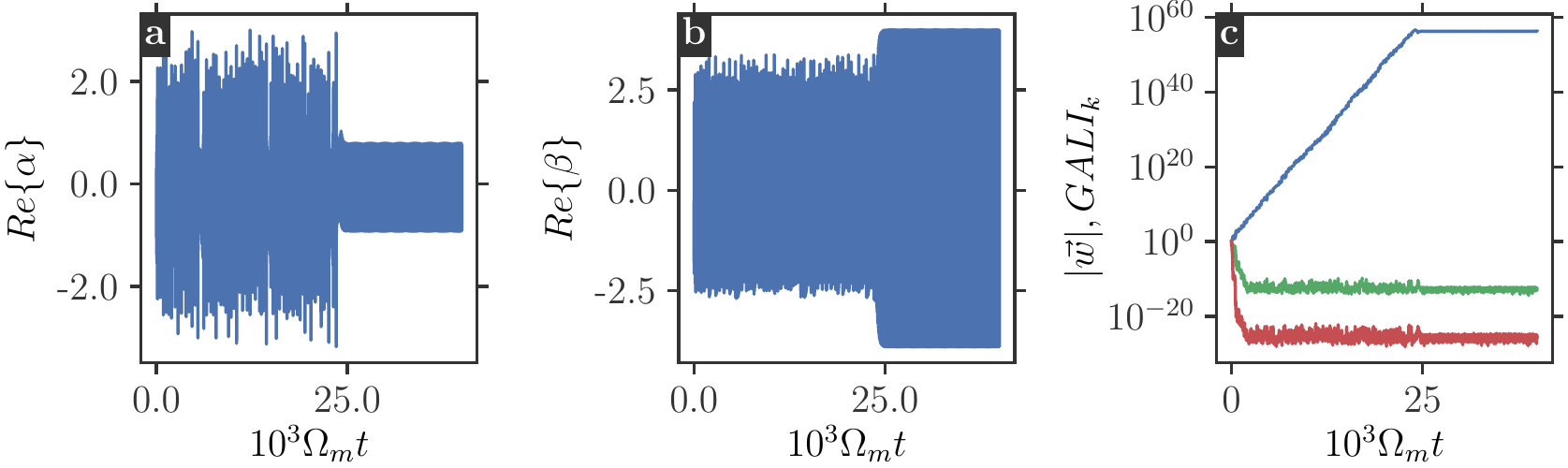}
	\caption{Classical transient chaos in an OM system. The dynamics of optical
	and mechanical degrees of freedom is shown in Panels~(a) and (b),
	respectively. In Panel~(c), we present the time evolution of three
	indicators, $|\vec{w}|$ (blue curve), the $GALI_{2}$ (green curve),
	and the $GALI_{3}$ (red curve). The parameters are $\Delta=0.5696\Omega_{m}$
	and $P=0.3$. All three plots clearly display the time instant at
	which the trajectory leaves the chaotic attractor and is attracted
	to the regular one. The fast decay of the $GALI_{2,3}$ is cut at
	values $O(10^{-12})$ and $O(10^{-25})$, respectively, because of
	the numerical precision of our method.}
	\label{transient_chaos} 
\end{figure}

Transient chaos in OM systems has been discussed for the first time
in Ref.~\cite{wang2016}. The authors of this paper argue that transient
chaos underlies the breakdown of the quantum-classical correspondence
in strongly dissipative OM systems, which display chaotic evolution
in the classical regime and regular dynamics in the quantum one. To
the best of our knowledge, the purely classical OM chaos has not yet
been studied. We explore it in the weakly dissipative case. A representative
example of transient chaos in classical OM is shown in Fig.~\ref{transient_chaos}.
The time evolution of the optical and mechanical variables, Figs.~\ref{transient_chaos}(a,b),
clearly manifests a crossover from the initially stochastic dynamics
to subsequent regular motion. The crossover is obvious also in the
behaviour of the modulus of the deviation vector $|\vec{w}|$, and
the $GALI_{2,3}$, Fig.~\ref{transient_chaos}(c). Before the crossover,
$|\vec{w}|$ increases while the $GALI_{2,3}$ decays exponentially,
confirming that the trajectory is chaotic. $|\vec{w}|$ stops increasing
at some time instant and oscillates around a nonzero value at longer
times. This means that the trajectory becomes regular. The decay of
the $GALI_{2,3}$ is cut at even much shorter times because of the
finite numerical precision of the method which has been used to solve
the equations of motion.

\begin{figure}[!ht]
	\centering
	\includegraphics[width=1\textwidth]{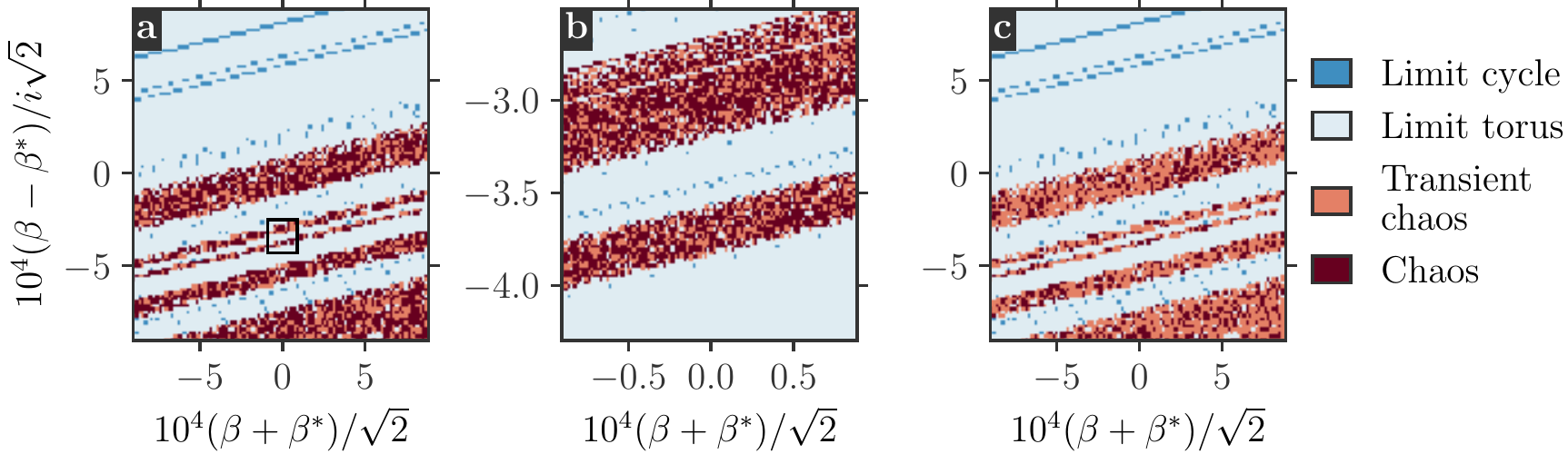}
	\caption{Panel~(a): Basin of attraction of attractors with different dimensions.
	The parameters are $P=0.395$ and $\Delta=0.61\Omega_{m}$. Panel~(b):
	Zoom of the area within the black box in Panel~(a). The basins of
	attraction display a similarly strong complexity at different scales
	(Panels~(a) and (b)). Panel~(c): The same basin of attraction as
	in Panel~(a) but with a longer realization time. The equations of
	motion were solved up to times $10^{5}/\Omega_{m}$ for Panel~(a)
	and $2\times10^{5}/\Omega_{m}$ for Panel~(c). Many pixels classified
	as chaotic in Panel~(a) are transiently chaotic with a large escape
	time.}
	\label{basin_attraction2} 
\end{figure}

The chaotic fractions of the phase space are elaborately intertwined
with the regions of transient chaos; see Fig.~\ref{basin_attraction2}.
We expect that this is a generic property, though details of the phase
space (whether a given pixel belong to the genuine or transient chaos)
are certainly sensitive to the cutoff time used in the empirical criterion
explained above. Fig.~\ref{basin_attraction2}(c) shows the same
basin of attraction as that drawn in Fig.~\ref{basin_attraction2}(a),
but now with the doubled cutoff $T'_{cutoff} = 2 \times 10^{5}\Omega_{m}$.
We note that many pixels, which were classified as chaotic in Fig.~\ref{basin_attraction2}(a),
are now classified as transiently chaotic in Fig.~\ref{basin_attraction2}(c).
Thus, many chaotic trajectories are actually transiently chaotic,
but with a large escape time $t_{{\rm esc}}$.

The high complexity of the phase space results in hypersensitivity
of the dynamics to the initial conditions. We have discussed this
phenomenon already in Sect.~\ref{HyperSens}; see Fig.~\ref{attractors}.
Fig.~\ref{basin_attraction2} suggests that the hypersensitivity
is generic in weakly dissipative OM systems which possess multistability
(co-existence of different attractors).

\section{Experimental relevance of our results \label{ExpRel}}

There are several experimental works devoted to OM systems that report
dissipation constants similar to (or even smaller than) those we have
used for our numerical simulations. Weakly dissipative OM resonators
can be fabricated in microwave systems \cite{rocheleau2010,teufel2011,yuan2015},
microresonators \cite{park2009,verhagen2012} and photonic crystals
\cite{chan2011,burek2016}, to name just a few platforms. The detuning
can usually be changed in a broad range. More important for investigations
of the nonlinear effects is the accessible range of the driving strength,
which governs the values of the parameter $P$. $P$ itself is not
convenient to describe the experiments, and it is better to consider the 
\textit{maximum} cooperativity $\tilde{C}$.
The value $P\sim0.1$ corresponds to a \textit{maximum} cooperativity 
$\tilde{C}\sim10^{6}$. This value agrees, for example, with the experimental
value reported in Ref.~\cite{peterson2019}. We thus believe
that the nonlinear phenomena described in the current paper can be
explored in the near future in modern experiments. In particular, phenomena 
similar to the Neimark-Sacker bifurcation have been already observed 
experimentally in Refs.~\cite{bagheri2013, seitner2017}.

We note also that some platforms have dissipation constants substantially
smaller than the values chosen for our study \cite{teufel2011, yuan2015}. We
have not considered such a weak dissipation but we think that nontrivial
nonlinear phenomena could be found in less dissipative OM samples
for substantially smaller values of $P$.

\section{Conclusions \label{Concl}}

We have demonstrated that the classical nonlinear dynamics of an optomechanical
resonator shows a great variety of nontrivial properties when the
dissipation is weak. This regime 
had not received proper attention in the few previous studies dedicated
to nonlinear OM dynamics, though it is of great experimental significance.

The phase space of the simplest OM system is four dimensional
and includes two mechanical and two optical variables. High dimensionality
and the presence of dissipation bring an extreme level of complexity
to any systematic study. This is because analytical methods are basically
unavailable while standard numerical approaches converge rather slowly.
To overcome these technical difficulties, we have suggested a novel
application of the $GALI$ method, which was initially developed for
Hamiltonian systems, to study attractors of the dissipative nonlinear
OM system. Our approach has several advantages. Firstly, it has proved
to be substantially faster than that based on an analysis of the maximal
Lyapunov exponent. Even more importantly for our goals, it allows
one to easily distinguish attractors of different dimensionality.

We have shown that weak dissipation strongly facilitates various nonlinear
OM effects, which can appear at substantially lower laser power as
compared to the previously studied strongly dissipative OM dynamics.
In particular, weakly dissipative dynamics becomes chaotic at $P\approx0.1$
(see the definition in Sect.~\ref{WeaklyDissOM}), one order of magnitude
smaller than the typical values of $P$ needed for chaos in the strongly
dissipative case.

Our choice of parameters has allowed us to reveal multistability,
i.e. the co-existence of different attractors. Their basins of attraction
are very complex and entangled. As a result, a tiny variation of the
initial conditions can completely change the dynamics on long time
scales, since the trajectory is driven to a different attractor. Such
a hypersensitivity to the initial conditions occurs even when the
dynamics is regular and there are no chaotic attractors. We believe
this to be a generic property of weakly dissipative OM systems.

Another generic feature reported in the current paper is the existence
of quasiperiodic attractors, or 2-dimensional tori, in the OM phase
space. We have investigated the transition from limit cycles to quasiperiodic
orbits, which, in turn, can undergo a transition to chaos. The latter
transition has some similarities to the well known period doubling
cascade and provides a new route to chaos for OM systems. Finally,
we have detected transient chaos. To the best of our knowledge, transient
chaos has not been observed in previous studies of classical OM dynamics.

In spite of the great power of our numerical approach, we have not
been able to obtain completely exhaustive information about weakly
dissipative OM dynamics. This is because scanning all possible combinations
of the four dimensionless parameters (rescaled power, detuning, mechanical
and optical dissipation) and a broader range of the initial conditions
is simply not feasible. We have focussed on exploring the phase diagram
in terms of power and detuning, while keeping the dissipation values
fixed. Thus, any complementary analytical method could be of great
importance. We believe that an extension of the method suggested in
Refs.~\cite{soskin_2008,soskin_2010,soskin_2012} might help to achieve
further progress. This method is based on the analysis of hyperbolic
trajectories in phase space. It has initially been developed for ac
driven dissipationless dynamics. However, its generalization to the
weakly dissipative case seems to be possible and promising.

We have argued that all the nonlinear OM phenomena which we have described
are within the reach of state-of-the-art experiments in optomechanics.
Moreover, it would be interesting to extend the present analysis to
OM arrays, which are known to have a tendency towards complex and
chaotic motion \cite{heinrich2011}. This could lead to exploring the complex
interplay of the Anderson localization physics, first predicted in Ref.~\cite{roque2017},
and nonlinear OM dynamics.

\section*{Acknowledgements}

We are grateful to Stanislav Soskin for useful discussions. This project has
received funding from the European Unions Horizon 2020 research and 
innovation programme under grant agreement No 732894 (Hybrid 
Optomechanical Technologies) and from the S{\~a}o Paulo Research
Foundation (FAPESP; process No 2012/10476-0).

\bibliographystyle{ieeetr}
\bibliography{Chaos-OM}

\end{document}